\documentclass[sigconf]{acmart}

\AtBeginDocument{%
  }

\copyrightyear{2026}
\acmYear{2026}
\setcopyright{cc}
\setcctype{by-nc-nd}
\acmConference[CHI '26]{Proceedings of the 2026 CHI Conference on Human Factors in Computing Systems}{April 13--17, 2026}{Barcelona, Spain}
\acmBooktitle{Proceedings of the 2026 CHI Conference on Human Factors in Computing Systems (CHI '26), April 13--17, 2026, Barcelona, Spain}
\acmPrice{}
\acmDOI{10.1145/3772318.3790584}
\acmISBN{979-8-4007-2278-3/2026/04}

\begin{document}

\title{Understanding Educators' Perceptions of AI-generated Non-consensual Intimate Imagery}

\author{Tongxin Li}
\email{tl434@njit.edu}
\orcid{0009-0002-8633-7394}
\affiliation{%
  \institution{New Jersey Institute of Technology}
  \city{Newark}
  \state{New Jersey}
  \country{USA}
}

\author{Katelyn M Reyes}
\email{kmr33@njit.edu}
\orcid{0009-0000-3452-0969}
\affiliation{%
  \institution{New Jersey Institute of Technology}
  \city{Newark}
  \state{New Jersey}
  \country{USA}
}

\author{Liezeil Jimenez}
\email{llj@njit.edu}
\orcid{0009-0007-7724-4811}
\affiliation{%
  \institution{New Jersey Institute of Technology}
  \city{Newark}
  \state{New Jersey}
  \country{USA}
}

\author{Katie S Nam}
\email{katie.s.nam@gmail.com}
\orcid{0009-0006-9951-9217}
\affiliation{%
  \institution{New Jersey Institute of Technology}
  \city{Newark}
  \state{New Jersey}
  \country{USA}
}

\author{Donghee Yvette Wohn}
\email{yvettewohn@gmail.com}
\orcid{0000-0001-5583-4430}
\affiliation{%
  \institution{New Jersey Institute of Technology}
  \city{Newark}
  \state{New Jersey}
  \country{USA}
}

\renewcommand{\shortauthors}{Tongxin Li et al.}

\begin{abstract}

AI-generated non-consensual intimate imagery (AIG-NCII) is an emerging social problem due to the advancement of AI tools. While recent incidents in middle and high schools have highlighted the urgency of this issue, there is limited understanding of what concrete supports schools need to effectively address AIG-NCII. To fill this gap, we conducted an interview study with 20 educators in the U.S. and investigated their attitudes, experiences, and practices related to AIG-NCII. Educators expressed concerns about both students' and their own vulnerability, as AIG-NCII may cause moral decline among students, while educators themselves could become victims. Nevertheless, existing practices in schools are limited, and they lack both training and systematic policies. Challenges such as a lack of resources, unclear legal boundaries, and limited knowledge of AI make implementation difficult. The findings of this paper contribute to interactive educational tool design, curriculum design, and policy-making, especially regarding the need for multi-stakeholder strategies to address issues surrounding AIG-NCII.

\end{abstract}

\ccsdesc[500]{Human-centered computing~Empirical studies in HCI}

\keywords{AI-generated non-consensual intimate imagery, generative AI, AI-generated image, deepfake, user-generated content, AI in education}

\maketitle

\section{Introduction}

Generative AI has advanced in recent years, proving adept in its ability to produce realistic imagery difficult to distinguish from imagery made by people \cite{partadiredja2020ai}. Openly available generative AI models such as DALL-E \cite{DALLE}, Midjourney \cite{Midjourney}, and Stable Diffusion \cite{StableDiffusion} not only allow people to experiment, but make it accessible to the general public to create highly realistic imagery \cite{bansal2024revolutionizing}. While such accessibility has led to positive innovations in education \cite{harry2023role}, creativity \cite{vinchon2023artificial}, and productivity \cite{perera2025sky, qiao2025use}, it has also introduced serious risks of misuse. Among cases of misuse, AI-generated non-consensual intimate imagery (AIG-NCII) has become one of the most concerning \cite{brigham2024violation, dunn2021women, farid2022creating, okolie2023artificial}. AIG-NCII refers to the non-consensual creation of explicit or intimate imagery produced with generative AI using someone's likeness \cite{brigham2024violation}. Abuse of AIG-NCII has the potential for psychological and social consequences for the affected individuals \cite{langlois_economies_2017, viola_designed_2023}. This issue is especially relevant in educational settings, where students represent both vulnerable victims and perpetrators \cite{noauthor_2024_nodate, CDT, noauthor_beverly_nodate, noauthor_opinion_nodate}. Several notable incidents across schools in the U.S, including cases in California \cite{noauthor_beverly_nodate} and New Jersey \cite{noauthor_2024_nodate, noauthor_opinion_nodate}, have drawn public awareness to the seriousness of the issue. Correspondingly, a survey conducted in 2024 highlights the increasing prevalence of AIG-NCII in schools across the U.S., yet schools remain unclear in their response policies and procedures \cite{CDT}. While previous research on AIG-NCII examined public perceptions \cite{kugler2021deepfake, umbach2024non} in addition to technical and legal frameworks \cite{anderljung2024protecting, kugler2021deepfake, chesney2019deep, citron2014criminalizing} as a society, there remains a gap in qualitative research analyzing AIG-NCII in schools, specifically educators' perceptions. 

Educators have a vital role in facilitating knowledge, understanding, and safety in students, directly engaging and spending considerable time with them \cite{de2019teacher}. Hence, educators are frequently the first adults to notice and respond to incidents among their students \cite{office2021protecting, mcdonough2024teachers}, including AIG-NCII. While prior work has examined K-12 educators’ perspectives on this issue, largely through a privacy lens \cite{wei2025we}, there remains limited understanding of the concrete infrastructural, procedural, and institutional supports needed to effectively address AIG-NCII in educational settings. Addressing this gap is essential to surface the practical challenges schools and educators face and to inform viable prevention and response efforts.

To explore this topic, we conducted an interview study with 20 educators across the United States, including K-12 teachers, school counselors, school psychologists, and higher education instructors, to learn about their attitudes toward AIG-NCII, existing practices in their institutions, needed practices, and challenges in implementing those practices. Our inquiry was driven by the following research questions:

\textbf{RQ1: What concerns do educators express regarding AIG-NCII in educational contexts?}

\textbf{RQ2: What practices are currently in place to address AIG-NCII?}

\textbf{RQ3: What measures do educators believe should be implemented to address AIG-NCII and how to implement them?}

\textbf{RQ4: What challenges arise in implementing measures to address AIG-NCII?}

We found that educators widely viewed AIG-NCII as a serious and growing problem in schools. Many educators had experienced these incidents directly or indirectly. However, few educators worked in schools with clear policies or training to handle such cases. Educators expressed concerns for students and themselves, for the power of AI technology, and for AI's ethical ambiguity. Educators emphasized the need for interventions at multiple levels, calling for specialized training about AI for educators, education for students on consent and AI ethics, and stronger collaboration with parents. They also advocated for better support from technology companies and more robust government regulation.

This paper contributes to ongoing discussions in Human-Computer Interaction (HCI) regarding ethical and social issues related to the use of generative AI from the perspective of the educators, who are unique stakeholders, because they are usually not the main perpetrator or victim, but are in a position to witness and potentially mitigate these problems. Educators have identified several problems that require action to address AIG-NCII, but these issues are complex and demand collaboration among multiple stakeholders. This research amplifies the voices of educators to expand current work on AI misuse, informs future interactive educational material design, and identifies what needs to be done in the school system.

\section{Background}

In this section, we provide a brief summary of related work on (1) the definitions and impacts of AIG-NCII, (2) the effectiveness of current practices used to address AIG-NCII, and (3) educators' perceptions of AIG-NCII.

\subsection{AI-generated Non-consensual Intimate Imagery (AIG-NCII)}
In this study, intimate imagery refers to "images and videos of people who are naked, showing their genitals, engaging in sexual activity or poses, or wearing underwear in compromising positions"\cite{StopNCII.Org}. AIG-NCII refers to the non-consensual creation, distribution, or threats made with intimate images produced using generative AI technology \cite{brigham2024violation}. This term is used in our paper to emphasize the non-consensual aspect of imagery specifically produced using generative AI technology such as face-swapping, lip-syncing, text-to-video systems, and voice conversion \cite{brigham2024violation}. This contrasts with the colloquially used term of "deepfake," which allows for a broader interpretation of consent and encompasses various methods of image production, such as photo editing software. Furthermore, the term "deepfake" does not capture the increased facilitation of imagery due to public accessibility of generative AI technology. Likewise, "pornography" may limit the range of what an individual considers to be intimate imagery \cite{mcglynn_beyond_2017}.

Repercussions of non-consensual intimate imagery (NCII) could be severe mental health consequences, such as depression, anxiety, post-traumatic stress disorder, and suicidal ideation \cite{dekker_zur_2021, frankel2018sexting, mcglynn2021s}. NCII also has social repercussions, including bullying, harassment, reputational damage, and social isolation arising from victim-blaming attitudes \cite{dekker_zur_2021, brigham2024violation, victimblaming}. Negative social repercussions and poor mental health can increase self-blame \cite{Meehan, FelipaSchmidt} and may prevent victims from seeking support \cite{victimblaming}. Social media and the rise of generative AI increase the prevalence of NCII \cite{FelipaSchmidt}, and while AIG-NCII may depict fake content using a person's likeness, the negative impacts on victims remain the same as other forms of NCII \cite{langlois_economies_2017, viola_designed_2023}. 


The impact of AIG-NCII is ubiquitous, affecting both ordinary people and prominent figures \cite{okolie2023artificial}. In South Korea, women have been the victims of a growing AI-generated sexual imagery epidemic, with reported instances reaching 297 as of July 2024 \cite{noauthor_south_2024}. In January 2024, AI-generated sexually explicit images of pop star Taylor Swift were posted on X, garnering over 45 Million views before being taken down 17 hours later \cite{noauthor_taylor_nodate}. Similarly, sexually explicit AI-generated videos of rapper Megan Thee Stallion were spread online in June 2024 \cite{martone_victims_2024}.

Although numerous papers address the increasing weaponization of AI technology against women or gender-diverse persons of different standings in the public \cite{noauthor_south_2024, noauthor_taylor_nodate, martone_victims_2024}, we are specifically interested in the impact of AIG-NCII affecting educational institutions across the United States. The rise of AIG-NCII poses significant challenges in education, where students increasingly fall victim to such abuse. Over the past three years, students' photos have been used to create AIG-NCII in middle and high schools across multiple states \cite{singer2024teen, Chavez_2023, Curi_Rebik_2024, González_2024, Haskell_2024, Haskins_2024, Tenbarge_2024, Johnson_2024, Liou_2024, Moseley_2024, Ciavaglia_2025, Hanna_2024}. One such scenario involves male classmates using AI to generate fake nude images of underage high school girls in New Jersey \cite{McNicholas_2023, noauthor_opinion_nodate, noauthor_2024_nodate}. Similarly, at a middle school in California, five students created and circulated AI-generated fake nudes of 16 eighth-grade students and were subsequently expelled \cite{noauthor_beverly_nodate}. In the 2023-2024 school year alone, both students and teachers reported significant amounts of AIG-NCII, depicting individuals from their schools, with the primary victims and perpetrators being students \cite{CDT}. 

Despite the proliferation of AIG-NCII in schools, over half of teachers in a survey conducted by the Center for Democracy \& Technology reported their schools failed to provide policies and procedures to teachers and students and have not responded to the increasing occurrences \cite{CDT}. Correspondingly, previous studies have emphasized the need to educate the public about AIG-NCII and to promote responsible data-sharing practices \cite{okolie2023artificial, RINGROSE2022102615, viola_designed_2023, wei2025we}. 

Although there are differences in research methods, studies consistently reveal that the public strongly condemns AIG-NCII \cite{fido2022celebrity, kugler2021deepfake, umbach2024non, brigham2024violation}. Two studies \cite{kugler2021deepfake, umbach2024non} particularly emphasized this consensus, with participants from representative U.S. samples and international populations expressing profound disapproval of deepfake creation and dissemination. The studies unanimously demonstrated that the public perceives such content as inherently harmful, regardless of the specific context or implementation. While the core perception of AIG-NCII remained negative, studies revealed nuanced variations in public judgments. One study \cite{fido2022celebrity} found that perceptual leniency varied with victim characteristics, such as celebrity status, gender, and context of image creation. Another study \cite{brigham2024violation} further elaborated on these variations, showing how judgment differs based on the relationship between the creator and victim, as well as the respondent's demographic and attitudinal factors. 

\subsection{Current Practices to Address AIG-NCII}

While several interventions have been proposed to address the misuse of AI, particularly AIG-NCII, their effectiveness remains uncertain, and most emphasize technical solutions rather than educational approaches. For example, the AI Misuse Chain \cite{anderljung2024protecting} conceptualizes AI misuse prevention through a three-stage framework: before misuse, during misuse, and after misuse. The proposed interventions are organized around these three stages. In terms of AIG-NCII, before misuse, there are two aspects that need to be considered: mitigating AI development and restricting resource allocation to prevent potential unethical applications \cite{anderljung2024protecting}. The authors recommended that developers train models to be less effective on misuse-relevant tasks, e.g. AIG-NCII. During misuse, they suggest mitigating harm by identifying and labeling AI-generated content or making it more difficult to automatically post such content. 
However, a study \cite{kugler2021deepfake} demonstrated that labeling such content as fictional did not mitigate its perceived wrongfulness, a finding reinforced by other studies that highlighted the severe moral condemnation of sexually explicit non-consensual synthetic imagery. After misuse, the studies \cite{anderljung2024protecting, kugler2021deepfake} discussed the introduction of laws and voluntary frameworks to govern the production and distribution of non-consensual deepfake pornography. 
Despite growing concerns about AIG-NCII, scholarly research has predominantly focused on technical solutions rather than educational strategies.
In this study, we examine existing practices and policies within the schools and educational frameworks to assess the extent of current preparedness and identify gaps in existing guidelines.

\subsection{Educators' Perception of AIG-NCII}

Existing research provides valuable insights into public perceptions of AIG-NCII \cite{fido2022celebrity, kugler2021deepfake, umbach2024non, brigham2024violation} and students' perceptions \cite{kopecky2025phenomenon}, but a critical gap remains in understanding the perspectives of educators. Unlike the general public, educators have direct responsibilities in guiding and supporting adolescents, a group particularly vulnerable to AIG-NCII \cite{de2019teacher, office2021protecting, mcdonough2024teachers}. Educators are on the front lines of education and responsible for delivering knowledge, shaping students’ understanding, and often serving as the first adults to notice, respond to, or be asked for help in such incidents \cite{de2019teacher, office2021protecting, mcdonough2024teachers}. 
Despite growing awareness of AIG-NCII, discourse on AI education has primarily focused on the formal use of AI \cite{uygun2024teachers, amado2024exploring, thararattanasuwan2024exploring, kaplan2023generative, jose2024educators}, such as integrating AI tools into teaching and learning experiences \cite{kaplan2023generative, amado2024exploring}. In contrast, students' informal AI use has received far less scholarly attention. A recent study published by Wei et al. \cite{wei2025we} investigated teachers' perceptions of what motivates students to create AIG-NCII and what potential interventions might be effective, but infrastructural dimensions remain underexamined. Our study builds on prior work by examining infrastructural, procedural, and systemic considerations within this informal AI use context.

While previous research has examined public perceptions of AIG-NCII \cite{fido2022celebrity, kugler2021deepfake, umbach2024non, brigham2024violation} and explored technical and legal frameworks for addressing this issue \cite{anderljung2024protecting, kugler2021deepfake, chesney2019deep, citron2014criminalizing}, a critical gap remains in understanding the infrastructural, procedural, and institutional dimensions of how educational settings handle AIG-NCII. By exploring educators' firsthand accounts of AIG-NCII incidents, current institutional responses, and barriers to effective intervention, our research addresses this gap. This work amplifies educators' voices as unique stakeholders who witness these harms directly and identifies concrete measures needed across multiple levels, from classroom curriculum to government policy, to protect students in educational contexts.

\section{Methods}

\subsection{Recruitment and Participants}

We conducted interviews with 20 educators in the United States to understand their perspectives on AIG-NCII. We recruited participants by posting recruitment information in two education-related Facebook groups and one Reddit group. Prior to posting, we obtained permission from group moderators and provided our Institutional Review Board (IRB) approval when requested.
Individuals interested in participating first completed a screener survey, where we collected demographic information (age, gender, race, occupation, job length, state, and sexual orientation). To be eligible for the study, participants were required to be currently working in an education-related role in the U.S. We aimed to recruit a diverse sample across gender, race, geographic location, and years of teaching experience.

We received 385 survey responses. For those who met our recruitment criteria we sent emails and invited them to schedule interviews with us. We did not find meaningful differences based on sexual orientation and therefore did not include it in the reported results. 
Of the 20 educators we interviewed, 5 were recruited through personal networks and 15 through posts in Facebook and Reddit groups. Participants ranged in age from 25 to 63 (average = 33.16). The sample included 8 males, 11 females, and 1 non-binary participant, located across ten states. Their education-related work experience ranged from 1.5 to 40 years (average = 7.35). See Table 1 for more detailed participant demographics.

\begin{table*}[t]
\centering
\caption{Demographic and Professional Information of Participants}
\Description{This is a participant table with their demographic information, including gender, age, race, occupation, job length (yrs) and location.}
\small 
\renewcommand{\arraystretch}{1.15} 
\begin{tabular}{lccllcc}
\toprule
\textbf{Participant ID} & \textbf{Gender} & \textbf{Age} & \textbf{Race} & \textbf{Occupation} & \textbf{Job Length (yrs)} & \textbf{Location} \\
\midrule
P1  & Male   & 39 & Asian           & University Instructor            & 1.5 & New Jersey     \\
P2  & Male   & 30 & Asian           & University Instructor            & 1.5   & New Jersey     \\
P3  & Male   & 63  & White          & University Instructor            & 40  & New Jersey     \\
P4  & Female & 29  & Asian                   & Board Certified Behavior Analyst
  & 3 & Georgia \\
P5  & Male   & 44  & Asian                   & High School Physics Teacher    & 17  & New Jersey     \\
P6  & Female & 27  & Black                   & Assistant School Counselor     & 3   & California     \\
P7  & Female & 39  & Indigenous              & University Instructor          & 4   & New Jersey     \\
P8  & Male   & 28  & Black                   & High School Math Teacher       & 5   & New York       \\
P9  & Male   & 28  & N/A  & High School Guidance and Counselor            & 7   & California     \\
P10 & Female & 25  & Black                   & Teacher and Counselor             & 6   & New York       \\
P11 & Female & N/A & N/A  & High School Guidance and Counselor          & 4   & New York     \\
P12 & Non-binary & 28 & Black                & Economics Teacher              & 6   & Illinois       \\
P13 & Female & 37  & White                   & School Counselor               & 2   & Virginia       \\
P14 & Female & 30  & Black                   & High School STEM Teacher       & 5   & New York       \\
P15 & Female & 25  & Black                   & Middle School Teacher          & 5   & Florida        \\
P16 & Male   & 32  & Black                   & Social Studies Teacher         & 10  & North Dakota   \\
P17 & Male   & 24  & Black                   & High School Teacher            & 5   & Michigan       \\
P18 & Female & 33  & Black                   & Math Teacher                   & 10  & Florida        \\
P19 & Female & 36  & White                   & School Psychologist            & 6   & Pennsylvania   \\
P20 & Female & 33  & Black                   & High School English Teacher    & 6   & California     \\
\bottomrule
\end{tabular}
\end{table*}

\subsection{Interview Procedure}
We developed the semi-structured interview protocol based on our research questions. All interviews were conducted by two authors online via Zoom. We followed the same protocol with each participant, proceeding through all questions in the given order while also asking follow-up questions and revisiting topics as needed based on the information provided. 

At the start of each interview, we explained our study and made sure participants understood we would discuss AIG-NCII in schools, then asked about their current role, job responsibilities, and how long they had worked in their occupation. We followed by asking if they had heard of AIG-NCII cases and how they felt about them, as well as how such images might affect students, influence their understanding of digital ethics, and which groups might be more at risk. For educators and school staff, we also explored whether they teach AI ethics, are aware of any school rules on this topic, or receive any related training. Finally, we asked all participants about gaps they see in AI ethics education, how schools and parents should guide students, how schools should respond when problems occur, and what roles technology companies and the government should play. Interviews typically lasted about 1 hour, ranging from 40 minutes to 126 minutes.


\subsection{Data Analysis}
All interviews were audio-recorded, and we transcribed the recordings using Zoom's automatic transcription tool, then checked them manually. To analyze qualitative data, we used thematic analysis \cite{Braun2006} to study the interview data. Instead of calculating inter-rater reliability, we followed a consensus-focused approach that aligns with interpretive qualitative research. 

Analysis was done collaboratively by four authors following an iterative, inductive coding process. First, three authors independently read and coded a random subset of six or seven transcripts each (about 33\% of the dataset). This step helped us familiarize ourselves with the data and develop a preliminary understanding of participants' perspectives. During this phase, we used a shared document to track emerging codes and their definitions. After the initial coding, each author reviewed and coded subsets of the other two authors so that three authors became familiar with all interview data. We then gathered and compared our codes, discussing any discrepancies and refining our codebook. We created a matrix to map interview questions to research questions, documenting key quotes in a shared spreadsheet. Then we divided the coding work according to our research questions. Each author took primary responsibility for finalizing the codes related to specific research questions. We continued to cross-check each other's coding and discussed any new codes, resolved disagreements through detailed discussion, and ensured we applied codes consistently. 

After finishing all coding, the four authors worked together to analyze the codes and identify patterns and relationships. We grouped related codes into broader categories, then refined these into themes and subthemes through discussion. The final codebook has 143 codes in 21 themes. To maintain consistency with the qualitative approach of this study, we adopted a consistent terminology to communicate the frequency of major themes \cite{zhang2022usable, liu2024exploring}. Our frequency scale uses terms that range from "none" (0\%) through "a few" (0\%-25\%), "many" (25\%-50\%), "majority" (50\%-75\%), "most" (75\%-100\%), and "all" (100\%) \cite{liu2024exploring}. 

\subsection{Ethics}

This research was approved by IRB at our institution. Before each interview, participants reviewed the consent form. We obtained written or verbal informed consent from all participants and ensured consent was completed before proceeding with the interview and audio recording. Each participant received a \$20 Amazon gift card. Participants could skip any question or stop the interview at any time without any penalty and still receive compensation. 

During the interviews, we did not ask participants to share or show any images. We encouraged them to only share what they felt comfortable discussing. We were careful when asking about students and specific cases, and we did not ask for detailed or identifying information. After transcribing all interviews, we removed names and other identifying details about people and institutions. In the paper, we refer to participants using P1–P20. The audio recordings and transcripts were stored on secure, password-protected drives and only the research team had access to them.

\textbf{Positionality statement.}
Our research team has five female researchers based in the United States. Most of us have training in human-computer interaction. For more than half of our team members, English is their first language. Two authors came to the U.S. as international students many years ago. We have spent years living, studying, and working in the U.S. Most of us attended U.S. middle or high schools. Our positions influence the questions we ask and how we understand educators’ stories. We do not claim to represent all educators or students, especially those in non-U.S. contexts or in very different school systems.

\section{Results} 

Our results highlight key themes related to educators' concerns (RQ1, Section 4.1), current practices for addressing AIG-NCII (RQ2, Section 4.2), educators' proposed measures (RQ3, Section 4.3), and challenges in implementing those measures (RQ4, Section 4.4).

Among the 20 participants, 8 reported AIG-NCII cases in their schools, while 3 had never heard of such incidents. The others heard about cases involving students or celebrities through media. Most cases involved male students creating non-consensual intimate images of female students for bullying or reputation damage. P7 shared a case of a female student: "Her ex-boyfriend fit her picture into porn and circulated it on Discord. It really shattered her. The worst part is [that] she had no legal recourse. The university said it was off-campus conduct and the student didn't want her parents to know. We just tried to solve it quietly." P15 reported that explicit images were printed and pinned on the victim's back. A high school math teacher (P8) shared a case targeting a teacher: "It was a school in my district. Because his teacher gave him a bad score on an English test, he made a mock-up of his teacher in an NSFW (Not safe for work) format. It went viral, and the teacher had to be laid off. It went too far before anyone confessed." These cases reveal that AIG-NCII in schools may be far more common than reported in the news or media, highlighting the need for more systematic prevention efforts.

\subsection{RQ1: Educators' Concerns Regarding AIG-NCII}

Most educators expressed \textbf{strong concerns about students' wellbeing} (Section 4.1.1) when discussing AIG-NCII. They recognized that AIG-NCII affects students in multiple ways, including their \textbf{moral development, social relationships, emotional and mental wellbeing, and academic performance}. A few educators also expressed \textbf{concerns about their own role and vulnerability} (Section 4.1.2). They were also disappointed that students used AI technology in harmful ways instead of creative ones. Finally, educators described their \textbf{concerns about AI's power and ethical ambiguity} (Section 4.1.3). They felt shocked by how realistic AI-generated images looked, making it difficult to tell fake images from real ones. A few participants worried that their own students, family members, or even themselves could become victims. Many educators tried to imagine how victims would feel: the embarrassment, isolation, and emotional pain.

\subsubsection{Concerns about AIG-NCII's negative impacts on students.}


The majority shared concerns about how AIG-NCII can negatively affect students. Their responses focused on moral decline, reputation harm, emotional/mental wellbeing, and falling academic performance.

\textbf{Moral Decline.} 
A few participants expressed deep concern about the "desensitizing effects" (P8) of repeated exposure to AIG-NCII on young people's moral development. A central worry was that students encounter explicit material before they have developed the maturity to process it appropriately, with one participant warning that "AI just accelerates the process (of overexposure) and makes it far worse" (P8). This early and frequent exposure was seen as normalizing content that should be recognized as harmful, eroding students' ability to distinguish right from wrong.
Other participants described a dangerous pattern where repeated exposure leads to acceptance. When students see AI-generated explicit content again and again, they stop seeing it as problematic. One participant explained that students "would see that kind of content over and over and start to think it's commonplace and normal" (P5), which could remove their hesitation about creating even more extreme material. This desensitization was also connected to concerns about addictive behaviors. One participant used their personal experience with pornography to warn that creating intimate AI images could trigger similar addiction patterns (P14). Overall, participants believed this exposure seriously harms young people's developing understanding of healthy boundaries and appropriate behavior.

\textbf{Reputation Harm.} 
Many participants highlighted the severe social consequences victims face once AIG-NCII is shared. A primary concern was the damage to reputation and relationships. Victims would experience public shame and judgment from their peers. The situation is made worse by the difficulty victims face in defending themselves: people often refuse to believe the images are fake, leading to broken friendships and damaged social standing. As one participant explained, "Some of them don't even have the power to explain that it is not their actual picture. No one would ever trust you" (P15).
This social harm typically leads to significant behavioral changes, with affected students withdrawing from school and social activities. Participants emphasized that this isolation comes not just from embarrassment but from a loss of trust in their surroundings. When victims feel unsupported by peers and even teachers, they struggle to trust anyone around them. One participant described this dynamic: "There's that fear of judgment. They begin to lose trust in their peers, sometimes even in the teachers" (P20). This combination of social withdrawal and lost trust creates a cycle that leaves students increasingly isolated and vulnerable to further harm, including bullying. P17 further emphasized the broader consequences of these experiences, noting that such situations can create power imbalances because victims of AIG-NCII become more vulnerable. In addition, P3 pointed out that AIG-NCII content can reinforce harmful gender and racial stereotypes, making certain groups feel unfairly targeted or objectified. 

\textbf{Emotional and Mental Wellbeing.} 
School mental health professionals (P13, P19) described the emotional impact as profound and potentially escalating. Students experience feelings of violation and betrayal that manifest in observable behavioral changes, particularly social withdrawal and depressive symptoms. The harm extends beyond immediate distress to fundamentally undermine students' sense of safety and security. They emphasized that the emotional dimensions of AIG-NCII frequently progress into clinical mental health concerns, including diagnosed depression and other psychological disorders, underscoring the critical need for early intervention and trauma-informed care approaches.

\textbf{Academic Performance.} 
Many participants explained that the mental stress from AIG-NCII exposure directly harms students' academic performance. The psychological impact affects the mental reasoning of students, which would subsequently impact their academic grades (P10). Beyond just lower test scores, affected students struggle with basic classroom functioning. One participant noted that students' "performances often dip due to concentration issues or fear of returning to class" (P19). The mental pressure makes it difficult for students to focus during lessons, leading them to withdraw from class participation and perform poorly on assignments.
A few participants described how this academic decline connects to broader withdrawal from school life. Students stop engaging in class discussions and avoid school activities they previously enjoyed, with the mental impact affecting both their academic and social performance. In more serious cases, the deterioration can be severe and rapid. As one participant warned, students "can just fall off the wagon and keep going down" (P11), suggesting that without intervention, academic decline can accelerate into a downward spiral that becomes increasingly difficult to reverse.

\subsubsection{Concerns for educators' role and vulnerability.} 

A few educators expressed concern that if such an incident were to take place, they might have failed to educate their students the proper use of AI. P14 reflected, "[If that incident happens,] I don't think there should be any teacher who would actually feel good… It's actually out of context of what you're teaching your students." They worried they had not adequately prepared students to understand appropriate technology boundaries. One educator reflected that such an incident would make them question whether they had successfully connected with students in ways that prevent harmful decisions (P8). These reflections go a long way in proving that AIG-NCII incidents do not only bring up concerns about the welfare of students but also force educators to undertake a critical review of their teaching practices. Beyond worrying about their students, a few participants expressed anxiety about becoming targets of AIG-NCII themselves. One high school math teacher who encountered a real case involving a colleague in his district imagined the experience as deeply humiliating (P8). P7 described feeling personally unsafe and deeply uncomfortable after realizing how easily this technology could be misused: "I felt unsafe… you know if things like this can actually be done, it is a very bad thing for anyone to experience."

A few participants expressed concerns toward students who used AI in harmful ways. Instead of applying the technology for creative or educational purposes, some students weaponized it to hurt and objectify their peers, particularly female classmates. As P6 shared, "It was a little bit disappointing that these boys have gone this far, and instead of using it for something good, we're actually using it to… objectify the females and females as a whole." One participant explained that witnessing such incidents motivated them to proactively educate students: "I felt sad, especially it's happening in the education setting... That made me come with the idea that I have to make my own students know about the bad effects and the risk in the online space" (P12).

\subsubsection{Concerns about AI's power and ethical ambiguity.}

A few participants admitted they were impressed by AI's ability to generate highly realistic images. P1 and P2 noted surprisingly advanced content quality that heightened their awareness of the technology's power. One participant described how AI is "getting better" at creating images that closely match what people envision in their minds (P2). The realism has reached a point where images can no longer be trusted at face value, as one participant warned: "we are to the point that we cannot trust things just based on the image, because anyone can make a super realistic image" (P1). This participant worried about vulnerable populations who lack the digital literacy to question image authenticity. Without understanding that such realistic fakes are possible, people, especially older adults, would likely accept AI-generated images as genuine evidence, making them susceptible to manipulation and misinformation.

Many participants are confused about the legal and ethical boundaries of AIG-NCII. A physics teacher noted the difficulty of separating the moral problems of AIG-NCII from legal questions about what counts as a violation. He said, "You can't stop someone from creating whatever they want on AI, but how they use it is very blurry right now. There's freedom of speech, but at the same time, spreading these images, although they are fake, is considered child pornography. Is it considered slander or libel? You're kind of defaming and embarrassing someone. So I think it's morally wrong to create the image, although not illegal. But it is illegal to spread it around." (P5) It reveals that educators struggle with uncertainty over the legal and ethical boundaries of AI-generated content.

\subsection{RQ2: Current Practices in Addressing AIG-NCII} 

Many institutions have created guidelines to address AIG-NCII after an incident occurred. Other institutions are reviewing their current policies to address and prevent future occurrences. Among the 20 participants interviewed, 6 reported that their institutions already had guidelines in place regarding AI-generated intimate imagery. The guidelines currently in place include disciplinary actions, preventive measures, generative AI ethics curriculum, and reporting protocols. In contrast, 9 participants reported that their institutions had no current guidelines, and 5 were unsure whether any such policies existed. In this section, we described four current measures in schools for addressing AIG-NCII cases: \textbf{reporting protocols in institutions} (Section 4.2.1), \textbf{disciplinary actions for perpetrators} (Section 4.2.2), \textbf{preventative measures} (Section 4.2.3) and \textbf{teaching generative AI ethics} (Section 4.2.4).

\subsubsection{Reporting protocols in institutions.}

According to few participants, institutions are beginning to implement structured reporting guidelines to ensure that students can safely and confidentially report incidents of AIG-NCII. In multiple participants' institutions, students are encouraged to report violations to trusted teachers or other staff members, such as counselors. From there, staff bring the issue to the appropriate school authorities. 

In many cases, students are afraid to speak up and do not want others to know about their report. To address this, a few institutions have established anonymous reporting systems  or committees so students can safely inform staff about potential incidents. P20's institution has a safe phone number, managed by the school counselor and teachers, that students can call or message privately. P15's instution designed a reporting system to redirect messages from a channel directly to a teacher's private inbox. In addition, they use a separate secure platform accessible only to teachers and higher-level administrators, so other students cannot see the message or determine who shared it. This confidentiality is to essentially ensure that "the bully does not recognize that it's you who reported him and begins to try to oppress them or try to punish them" (P15). P6's institution uses online and physical forms to anonymously report incidents. P9's institution has created a student-led digital actions committee, overseen by teachers, to investigate incidents reported by students, so that student's can safely speak about this without judgment.

Once an AIG-NCII report is made, institutions follow established guidelines to properly assess incidents, determine disciplinary actions, and respond appropriately. P9's institution launches an investigation to uncover details, such as which parties were involved, before deciding what level of disciplinary action should be taken. P12 described how their institution's investigation process includes speaking with both the perpetrator and the victim to gather evidence and confirm that the incident was created by the perpetrator. Along with conducting investigations, a few institutions offer direct support for victims of AIG-NCII after an incident occurs. For instance, P19's institution follows a structured protocol where they immediately provide assistance for the victim. Another institution (P15) has created a dedicated support group for victims of AIG-NCII. 

\subsubsection{Disciplinary actions for AIG-NCII perpetrators.}

A few educators noted that their institutions have established clear disciplinary consequences for students who create or share AIG-NCII. These measures are used not only as guidelines for addressing violations, but also to discourage students from engaging in harmful behavior.

In response to incidents of AIG-NCII, a few institutions involve parents as part of the disciplinary process. For example, P19 explained that once a violation occurs, their institution promptly communicates the situation to the parents of the students involved. P12 shared that "the perpetrator has to face the consequences but before that… we make sure we call both parents of the victim involved and the perpetrator… so that they also would be involved in what decision the school is taking." In the event of a serious violation, some institutions reported the incident to law enforcement or authorities. P19 explained that the decision depends on "what the matter is, what the situation is at hand, and if there is a need to escalate it outside the school premises."

School suspension is another action taken to respond to AIG-NCII violations. At P12's institution, students are explicitly informed that engaging with AIG-NCII is completely unacceptable, and violators will face school suspension. By suspending students, this institution sends a clear message that this misconduct has real and immediate consequences. P8 added that in his institution, expulsion may be considered in extreme cases.

\subsubsection{Preventative measures against AIG-NCII.} 

To prevent future AI misuse and incidents of AIG-NCII, one institution is developing proactive defense guidelines. P15's institution has taken precautionary steps to create solutions for future purposes. She explained, "We've installed new security systems, so that things like this [AIG-NCII] will not just go through, even digitally. Things like this will not go through under our noses." Her institution also has a concern drop box where students can privately share their questions and concerns to be discussed in interactive sessions with educators. Additionally, the institution asked parents to monitor their children's phone activity to ensure they are not doing something that could cause harm later. These guidelines create a temporary system to prevent AIG-NCII cases from happening in the future.

\subsubsection{Teaching generative AI ethics.}

According to the interviews, a few institutions are beginning to include generative AI ethics in their curriculum. They are teaching students about the ethical versus unethical use of generative AI. In P20's institution, all teachers were advised to talk about generative AI ethics in their lectures and have conversations with students to spread awareness. She explained, "We talk more about consent. We try to let these students know how to identify it, know the possible threat and what to do in those circumstances. We teach them about the kind of things that they can put out there on social media, and the kind of things that they cannot." P12's institution teaches students that AI-generated intimate imagery is a violation and the "exponential risks" behind it. They warn them about peer pressure and harassment, and remind them that AI should be used for creative innovation, not harm. 

Other institutions do not have generative AI ethics specifically in their curriculum, but do teach digital ethics in their class. P6's institution has classes on digital ethics, where they teach about the ethical use of digital resources. They also have informal classes and one-on-one classes, where a dedicated department of counselors speaks with students about ethics and cyberbullying. P3, a university professor of 40 years, has added AI literacy into his syllabus, teaching students how to properly utilize AI tools. P16 teaches the ethical use of AI in his social studies course, emphasizing not to use AI to make others feel bad about themselves. 

\subsection{RQ3: Educators' Proposed Measures and Implementation Strategies to Address AIG-NCII} 

All participants suggested a variety of measures they believe should be implemented to mitigate AIG-NCII, along with recommendations on how to implement them. Participants who did not have guidelines in place often pointed to strategies already implemented in other institutions and suggested that similar approaches should be adopted in their own. Participants highlighted the importance of broader conceptual measures, such as \textbf{training for educators} (Section 4.3.1), \textbf{student education} (Section 4.3.2), and \textbf{parental involvement} (Section 4.3.3). Others emphasized infrastructure measures, including the \textbf{development of defined guidelines} (Section 4.3.4) and \textbf{clear reporting protocols} (Section 4.3.5). At a societal level, they suggested \textbf{professional and financial resources} (Section 4.3.6), \textbf{technical support from technology companies} (Section 4.3.7), and \textbf{stronger government laws} (Section 4.3.8). 

\subsubsection{Formal AI training for educators.}

Many participants believe that educators should receive more training in AI digital ethics, AI tools, and teaching students how to properly use AI. Out of the 20 participants interviewed, only one educator (P16) received formal training on the ethical use of AI. A few participants had attended informal AI training, voluntary lectures, or workshops. However, the majority of participants have not had training or are unaware of available training on AI ethics. P19 shared that educators should be trained on AI digital ethics to better teach their students. She stated, "You cannot teach what you don't know…"

Many participants expressed that educators should receive training in AI tools to develop a clearer understanding of their functions and possibilities. They found that educators lack knowledge about AI, making it difficult to guide students. P6 shared, "[educators] don't even know how far AI can go. They don't know the possibilities of children actually using this AI to create intimate images to this level, for example,... if the girls in my class did not bring it to my notice, I would not be aware that something like that can be done with AI." There is a large knowledge gap that many educators do not realize the power AI has, so they can not effectively help their students. P3 brought up that if educators do not know how to use or appropriately prompt AI, how can they correct it or guide students in the right direction? Therefore, a few participants suggested implementing educator training on AI tools. P11 believed there should be teachers' professional development where AI experts train educators on the use of AI and any future versions of AI, as well as online courses, workshops, and educational software and tools. P15 suggested technology workshops, where teachers physically use computers to learn AI, so they have the knowledge to detect AI content and take control of a situation.


A few participants also believed that educators should be trained how to handle AIG-NCII situations. P15 emphasized that teachers should learn to recognize and report digital abuse "discreetly and also compassionately." P2 highlighted the importance of "educating instructors on what's happening currently, and how students use AI to do their coursework. And what has happened in other states, like the intimate imagery issues..." P20 shared, "I think we should properly train the staff and every staff in the school. It's not just a school counselor problem or the English teacher's problem. Everybody should be aware of how to deal with this issue in a way that is trusting, safe, and probably discreet."

\subsubsection{AIG-NCII education for students.}

Many participants emphasized that students often lack awareness of the risks and harms associated with AIG-NCII, highlighting the need for early and explicit education on AI ethics and literacy. A few participants noted that students may not fully understand the emotional, social, and legal repercussions of AIG-NCII. Thus, many participants believed that education regarding AI intimate imagery should be incorporated into schools through a required subject in the curriculum. Within these required courses, participants recommended that students be taught online privacy, consent, digital safety, and the legal and emotional consequences of creating or sharing AIG-NCII content. They pointed out that there is insufficient focus on issues like consent and privacy, especially deepfakes and synthetic media, which are rarely discussed in schools. Participants called for more comprehensive teaching on these topics to strengthen students' understanding of AI ethics. 

P20 suggested setting up more workshops and bringing professionals to come and teach the students how to responsibly use AI as a tool. P15 said that educators should, “Encourage students through this curriculum to explore AI creatively like writing poems, coding games, generating arts in a positive and educational scene. Not manipulate it, to embarrass, to deceive someone, or to edit an image that is not yours without the owner's consent. Not only teach them what AI can do, teach them to use it to create better things.” In addition to this positive guidance, another participant highlighted the importance of how AI is framed for students. P2 stated, "I think positioning AI as something wrong or like something unethical is something we should not do. Educators, too, should see AI as more a companion and more like a tool… Don't restrict students at the same time, like hold hands for them."

The majority of participants stated that education should start when students are exposed to social media, technology, and other online platforms where AI-intimate images might appear. Therefore, participants determined that education should begin around the time of middle school and high school. P6 added that educators need engaging educational material, suggesting: "Maybe they can have resources like, maybe since children actually love colorful things, no matter the age they are, maybe some video, or maybe some animated way of teaching these children. Okay, these are the things you should watch out for. These are the things you should not do. These are the kind of images that are classified as intimate images." In order to guide students regarding the ethical use of AI-generated images, the majority of participants suggested that all teachers be responsible for teaching AI ethics and literacy, rather than specialized teachers alone.

\subsubsection{Parental involvement in guiding children's ethical AI use.}

When asked about parental involvement in guiding their children on AIG-NCII, most participants suggested that parents be involved in educating their children. Participants noted that students are likely to experiment more with technology at home. P18, for example, stated that children don’t use their phones in school and therefore suggested that parents play a larger role in monitoring their kids, as children are more likely to use their “gadgets” at home. Additionally, a few participants felt that they have little control over what children do at home compared to in school, making it important for educators to collaborate with parents to raise awareness. Therefore, parents should also play a role in educating their children on appropriate behavior regarding AI use. 

Many participants suggested that parents first educate themselves on AI and what their children may be exposed to and involved in online to better educate their children. Participants such as P19, a school psychologist, expressed concern that parents are not aware of issues with AIG-NCII or experienced enough with AI generative technology. To combat this, P15 recommended that parents understand the risks of AI and convey them to their children. They suggested family workshops and home newsletters with tips, warning signs, and age-appropriate language to learn about digital content. By gaining more knowledge on the topic, parents can teach their children about ethical AI use as well as the consequences of its misuse. Parents can deter their kids from abusing AI by having conversations about the impacts of their actions. P20 suggested parents discuss possible risks and correct responses before giving kids a phone. In addition, P4 suggested parents ask questions such as "What if this happened to you?" and  "How would you feel if an image was created about you without your consent?" to put their children in the shoes of the victim.

The majority of participants also stated that parents should monitor and restrict online activity. Parents are responsible for knowing what their kids are doing online to both prevent misuse and protect them from harm. A school psychologist (P19) noted that children do not know how exposed they are online. Therefore, it is the responsibility of the parent to be aware of their children's activity and restrict their online use. 

A few participants also suggested parents hold open conversations with their kids to foster "safe spaces" for communication. Open conversations between parents and children builds trust, making kids more likely to seek support if something happens. P20 stated, "we need to present a safe space, for, you know, so that when something like this happens, the children are more willing to confide in us than keep it to themselves." Additionally, when parents actively discuss these issues with their children, the information shared at school is reinforced at home. P20 added, "They [kids] are hearing the same information almost everywhere they go. So it's sort of registered in their brains. That there are penalties for doing this. You should not be caught doing that." 

\subsubsection{Defined guidelines on AI use and ethical boundaries.}

When asked, most participants stated that institutions today lack defined guidelines on ethical use of AI. They believe that clear guidelines and school policies should be implemented to help mitigate AIG-NCII. Participants emphasized the need for the guidelines to outline appropriate AI use and ethical boundaries, with defined consequences for violating them. 

Most participants called for guidelines that define the appropriate use of AI tools and the risks and responsibilities of using them. P4, who is a board-certified behavior analyst, believed that there should be guidelines specifying what students can use AI tools for, without mentioning AI intimate images, to prevent more people from engaging with AIG-NCII. She explained, "You don't even have to mention intimate images because as soon as you say, 'Don't do something,' the human brain immediately thinks about doing it... So if there's going to be a guideline on it, it needs to be phrased in a way of, 'This AI tool is meant for the following things.'" 

A few participants suggested that guidelines clearly state what students should and should not produce with AI. For example, P5 believes that there should not be anything explicit, sexual, drug-related, or death-related content generated on AI under any circumstances. P15 suggested classifying AI-generated non-consensual imagery as harassment, bullying, and digital or sexual misconduct to help students understand the consequences of the images. P13 believed that intimate images, or any image that aims to tease, mock, or spread a false image of a student, should be explicitly defined as bullying with clear punishments outlined in school guidelines. A few participants suggested that these guidelines and consequences should be explicitly communicated with students and parents, while P4 specifically urged they must agree to follow these policies via a contract. In addition, P15 suggested the consequences for creating, sharing, or threatening to share such images should be shared with incoming and continuing students and their family at the start of every school year.

\subsubsection{Reporting protocol for AIG-NCII incidents.}

Many participants urged for defined reporting protocols, with step-by-step procedures for responding to AIG-NCII incidents or ethical breaches. P2 emphasized the importance of having clear methods to track and hold individuals accountable. He highlighted the need for effective reporting tools and systems that ensure incidents can be verified beyond reasonable doubt. P15 suggested a school-wide response plan for incidents, outlining when to notify school authorities or law enforcement, procedures for collecting and handling evidence, and measures to support the victim, including a safety plan. P19 called for resources and support to manage cases. 

Many participants highlighted the importance of making response guidelines that support victims of an AIG-NCII incident. A few participants called for clear, survivor-centered reporting protocols and safe reporting channels. For example, P7 suggested confidential disclosure reporting options and support teams trained in trauma-informed care to help victims. Others advocated for a restorative process for victims. P15 suggested that response protocols should include how to support the victim, and provide them counseling and a safety plan. She shared that the guidelines should include "Steps to restore the affected student reputation... Also, to reintegrate them safely, without them feeling too disgruntled." P13 also explained that the guidelines should clearly state the process, who will be involved, and how things will be made right. The guidelines should also be available to each student, so everyone is aware how cases will be handled. 

\subsubsection{Professional and financial support for AI education.}

The majority of participants believed that schools need financial support and professional support from experts to better handle AI issues. Participants felt that access to external AI ethicists, legal experts, technical experts, and mental health professionals would help guide complex cases. They would aid in teaching educators and students, and handling AIG-NCII cases. Additional financial resources would support hiring experts. P18 felt that there needs to be more knowledgeable teachers in the AI field, thus funds are needed to employ more educators to properly teach students. Additionally, P18 and P10 believed that schools need funds to keep up to date with new technology and a supportive internet, so teachers are equipped to educate students. 

Many participants called for collaboration with professionals to manage AIG-NCII cases and funds for their help. P1 explained, "We just need someone to organize and manage those existing resources… We just need a designated group of people offices who are in charge of this issue. So in that sense regarding the resources, we just need money to hire those people, to have an office for those people, for them to keep working, and it just always costs money." P15 recommended collaborating with legal experts or law agencies to  create guidelines that align with local, state, and federal laws. P13 and P20 also felt that specialists, like therapists or psychiatrists, should be involved in the restorative process of victims. These professionals could also help, retrain, and better equip school counselors and other staff.  A few participants noted that employing these professionals would require more funding. 

\subsubsection{Technical support from technology companies.} 

Most participants believe that companies have a responsibility when it comes to AI intimate imagery. These responsibilities include preventing AI from creating inappropriate content and restricting minors from certain tools, employing reporting tools, and educating violators. 

Many participants felt that technology companies should implement measures to restrict misuse of their AI tools 
as these companies have a responsibility to prevent AI from generating inappropriate or harmful content, particularly AIG-NCII. They argued that AI systems should be trained to recognize and block sexually explicit or unethical prompts before producing any output to avoid irresponsible use from students. P1 and P6 pointed out that ChatGPT blocks certain outputs, so all technology companies should train their AI systems to detect inappropriate prompts, screen them out, and refuse to generate a response. In addition to refusing requests, P2 suggested that platforms should also explain why the content is inappropriate, helping to educate the young users on ethical boundaries. P12 mentioned that technology companies, like Google, should consistently remove deepfake creation tools, apps, and services from search engines.

Many participants highlighted the need for strict safeguards for certain age groups to help prevent the misuse of AI tools. Two participants (P8 and P13) pointed out that there are parental controls and content filters on social media or television, and these safeguards need to be implemented in AI tools to help limit children's exposure to harmful material. Others added that these companies should enforce minimum age requirements to access or use AI tools that could be misused. They emphasized that technology companies should develop AI tools with different age groups and ethics in consideration. A few participants highlighted the importance of reporting mechanisms and moderation tools to address AIG-NCII. They felt that technology companies should employ manual or automated moderation to review and remove AI-generated intimate or harmful images. P9 suggested that companies have the responsibility to stop the generation of harmful images at the source since "they are the one that started it from their own end, they can actually stop it." P17 added that reporting tools should be part of broader measures to restrict misuse, particularly among minors, in order to reduce harassment, retaliation, and exploitation. 

\subsubsection{Government regulation and legal frameworks for AIG-NCII.}

The majority of participants believe that the government should establish strong laws to address AIG-NCII. Many participants raised concerns about the current lack of enforcement. One noted (P2) that "there are very hardly any regulations these days… Government apparently does not care a lot, seems like, at least for now." He pointed to Europe's stricter GDPR model as an example of how stronger privacy and AI-related laws could empower users in the U.S. Participants suggested that the government should create strong and clear laws to protect people from the misuse of AI-generated intimate images. These laws can help stop harm, protect privacy, and define consequences for harmful actions.

A few participants believed that the government should hold people accountable for creating or spreading AIG-NCII. At the same time, P13 warned that without explicit laws, perpetrators slip through loopholes: "There was no law anywhere that catered to this issue… they kind of slipped out of it, and then it looked as though it's not that serious. So I do think that the government should actually make policies… and if people still abuse it, they would be punished." They also mentioned it is important for the government to protect minors. P5 explained, "For the protection of kids, the government should at least push for safeguards." Others suggested legal reforms to expand existing child protection laws to cover synthetic media. 

Additionally, a few participants believed the government should set guidelines for responsible AI use. P10 suggested "establishing guidelines for responsible AI development and use", while P12 emphasized that the government must "work with tech companies… and establish a law that prohibits the use of AI intimate imageries". Regulation of technology companies was seen as essential by a few participants. For example, P18 said "Big tech companies should be regulated so that some things can be removed, some things can be curtailed, and they'll have more control over it."

\subsection{RQ4: Challenges in Implementing Measures to Address AIG-NCII} 

Building on the educators' proposed measures, in this section, we report what challenges they may face when implementing those measures. We highlighted six key themes: \textbf{the policy-making process} (Section 4.4.1), \textbf{students' receptiveness} (Section 4.4.2), \textbf{educators' cooperation} (Section 4.4.3), \textbf{parents' collaboration} (Section 4.4.4), \textbf{the school board's attitudes} (Section 4.4.5), and \textbf{limited resources} (Section 4.4.6).

\subsubsection{Complicated policy-making and implementation process.} 

Many participants identified challenges in AIG-NCII policy design and implementation. The first major challenge is the rapid development of AI technology. Policies become outdated quickly because AI advances faster than institutions can update their guidelines. P1 explained that new AI capabilities can emerge within several months of publishing a policy, creating significant gaps in coverage. This problem is made worse by administrative delays: policies must be reviewed and approved by multiple levels of management before they can be implemented, which further postpones timely updates (P13). The second challenge involves unclear accountability mechanisms. P2 raised important questions about responsibility and tracking, asking who would be held accountable and how violations would be monitored. Most AIG-NCII behaviors fall into a gray area where ethical boundaries are not clearly defined. This ambiguity makes it difficult to establish appropriate consequences and measure policy compliance effectively. Enforcement proves difficult as individuals "try to exploit loopholes" (P15), making monitoring and enforcement challenges.

\subsubsection{Students' receptiveness and communication strategies.}

Many participants said it was hard to get students to take the issue seriously, making it challenging to ensure they followed the protocols. The nuanced dynamics between teachers and students were explained by P8: "No matter how cool a teacher is, he's still a teacher… the No.1 enemy of a student." While not all teachers and students are in an adversarial relationship, participants noted that for many adolescents, this perception can reduce receptiveness to guidance. A few participants observed that some students intentionally act against instructions: "Most of them also do what you ask them not to do" (P14). P4 warned that introducing AIG-NCII into sexual education might potentially spark curiosity to misuse. She elaborated, "I'm thinking that it would be a good idea to incorporate AI intimate imagery into some part of sexual education. However, it needs to be delivered in a much better way, rather than just briefly mentioning it and opening up a whole can of worms. Like, 'Oh, I didn't know AI could do this. Let me try.' Right? That might be worse. We do things out of good intentions, but they end up the wrong way." 

Educational measures taken against using AIG-NCII has the possibility of catalyzing the issue by introducing it to students. While participants call for the education of students on the issue, some worry that students may not be aware of the possibilities of AIG-NCII until it is mentioned by educators. The dynamics of teacher-student relationships and teenage behavior create a lot of difficulty when deciding how to correctly introduce and educate students on the issue. Hence, it is necessity to design the right curriculum that provides not only knowledge but trust and engagement. 

Communication style poses additional challenges. Even when educators know what to teach, effectively conveying the message remains a prominent challenge. P3 described rule enforcement as "a slippery slope," explaining: "You cannot legislate everything people need to do. People should realize that this is something I should or should not do. We should have our own codes, ethical codes." Fostering internal ethical awareness may be more effective than mandated policies. Gender also emerged as a barrier, with one male educator (P1) expressing uncertainty about discussing AIG-NCII with female students: "If I have to talk about intimate images to female students way younger than me, and I'm a middle-aged male... I don't know what the best way to teach that kind of topic is."

\subsubsection{Educators' knowledge, "buy-in" and engagement.}

Beyond communication with students, many participants highlighted that educators' limited understanding of AI capabilities, how AI-generated images work and how students can get access to those tools is a challenge in implementing measures. As P15 noted, "Some of our teachers are not actually tech-savvy enough to understand how AI could be used to fabricate images." Such views apparently indicate an urgent need for highly specialized training that would acquaint the educators with all the potential harms which accompany new technologies. Ensuring educators' active engagement also emerged as a challenge. One educator (P4) emphasized the need to gain teachers' "buy-in," cautioning that "if we start with the negative, like, 'This is something dangerous,' it will just become a whole ban of AI in schools." Another educator (P9) stressed that training should focus on building readiness and willingness to learn, noting "that should be the readiness for the teachers to actually want to learn." Without teacher interest and commitment, efforts to address AIG-NCII may not be effectively implemented. At the same time, participants pointed out that teachers already have substantial workloads and responsibilities. Such pressures may limit the time and energy teachers can dedicate to addressing AIG-NCII, even when they recognize its importance.

\subsubsection{Parents' collaboration.}

A few educators emphasized that cultivating children's digital literacy also requires the support of parents, who spend the most time with them. They are better positioned to communicate in age-appropriate ways. Some parents may complain when schools introduce guidelines on AI-generated intimate images. They may view such education as exposing their children to inappropriate content. School supervision alone is insufficient without parental involvement at home, but schools lack the authority: "We do not really have the... rights... to dictate to their parents." (P16) Without active parental support, efforts to address AIG-NCII may face resistance or be less effective.

\subsubsection{School board's attitudes.}

A few participants thought before officially approved guidelines are issued, school board perspectives can also pose potential challenges. The board review process is necessary but may face resistance from some board members, particularly if concerns arise about addressing such sensitive topics. However, proper implementation focused on student education and protection would likely gain approval (P7). P13 mentioned that multi-level approval systems present structural challenges: policies must go through all management levels, requiring roundtable discussions, drafting, and widespread availability. The greatest challenge is making this process as quick as possible while ensuring effective information dissemination.

\subsubsection{Limited resources.}

Beyond stakeholder collaboration, limited resources create implementation barriers. A few participants struggled to find materials helping students understand AIG-NCII consequences, and developing educational resources requires specialized professionals. Since this field is relatively new, few experts understand both the AIG-NCII issue and educational approaches to address it, creating broader societal challenges. Schools must decide whether to budget for combating potential technology-related damages, requiring funds and stakeholder coordination for hiring specialists and materials. Government legislation demands even broader expertise networks, including lawyers, youth education experts, psychologists, and sex education workers, further increasing costs for authorities.

\section{Discussion}

\subsection{Main Takeaways}

\subsubsection{Educators' concerns about student moral decline and their own vulnerability.}

Among the various concerns identified in our interviews, the most interesting finding is that educators expressed concerns about their students' moral decline. While previous work related to AIG-NCII has primarily focused on privacy \cite{wei2025we} and technical perspectives \cite{anderljung2024protecting, kugler2021deepfake}, participants in this study raised concerns about humanity. We also found that educators were concerned about potential harm to students' reputations, mental wellbeing, and academic performance, which is consistent with previous research on cyberbullying \cite{macaulay2018perceptions, baraldsnes2015prevalence, takizawa2014adult, li2008cyberbullying} and NCII's impact on adults \cite{brigham2024violation, dekker_zur_2021, victimblaming, frankel2018sexting, mcglynn2021s}. However, an important difference emerged: unlike online harassment without AI technology, which typically involved individuals harassing others with words created by the perpetrators themselves through digital platforms, AI technology now serves not only as a platform but also as an active tool for creating harassing content. With the advancement of AI technology, the issue of AIG-NCII has become more concerning than cyberbullying because AI is directly being used to create realistic fake images as a powerful "weapon" to harass or bully other people.

Beyond concerns about their students, educators also feared that they could be targeted themselves. Many participants said they felt unsafe or uncomfortable because AI can easily create realistic fake images. They worried that students or others might use AI to make harmful images of teachers, which could damage their professional reputations or personal wellbeing. This concern has not appeared strongly in previous cyberbullying research \cite{macaulay2018perceptions, baraldsnes2015prevalence, takizawa2014adult, li2008cyberbullying}, where educators are mainly seen as helpers, not possible victims. This dual vulnerability has not been highlighted in prior work on cyberbullying or AIG-NCII, and it suggests that future interventions should consider protecting educators as well as students. It also underscores growing uncertainty about AI’s ethical boundaries, indicating that the risks of AIG-NCII extend beyond student interactions and may affect the broader school community.

\subsubsection{Educators are calling for formal AI training for students, educators, and parents.}

Educators emphasized an important and pressing need for comprehensive AI education. Most educators had not received training related to AI ethics or AI-generated harms, which limited their capacity to advise students and respond appropriately. They believed teachers needed proper training and access to experts who could teach them about AI's capabilities, limitations, and how to intervene appropriately. At the same time, they think students should be taught regarding consent, behavior, and the emotional and legal aftereffects of misuse. They believed that it would be important for such learning to be incorporated into the standard curriculum and not merely delivered as one-off sessions.

Prior research has shown the importance of AI education among educators \cite{daher2025integrating}. Nevertheless, members of the educational community have noted important considerations regarding the structure of such educational programs: AI training often lacks connection to theoretical pedagogical frameworks and fails to incorporate sufficient critical reflection on the challenges and risks of AI implementation \cite{zawacki2019systematic}. Recent work to establish models of AI educational programming for educators further establishes the importance of covering areas of equity, ethics, and culturally responsive teaching to enable educators to successfully instruct the K-12 community \cite{black2024framework}. Our findings align with these concerns. Although educators recognize the importance of such educational needs, AI educational programming cannot merely contain technical aspects; it should also involve pedagogical considerations and important areas such as ethics and cultural.

Participants pointed out the involvement of parents, as many children interact with technology only within the home environment. As such, it was recommended that the school provide an opportunity for parents to learn information that would guide them on the proper use of generative AI tools, as they are often unfamiliar regarding the potential misuse thereof. The above highlights the various insights that were gleaned, and how the use of AI requires consideration and utilization of the respective tool, as it stands as one of the important levers that must be understood regarding the creation of safe and informed virtual environments.

\subsubsection{Current practices in school to address AIG-NCII are lacking standard procedures.}

Our findings reveal that the majority of schools lack proper and standardized processes for dealing with occurrences of AIG-NCII. A few of the participants indicated that their respective institutions had some form of protocol, which was created after the incident had taken place. The educators had to make decisions without clear school procedures for what happens after they report incidents, during the investigation, and for the victim's wellbeing. All the above suggest that there are no proper standardized processes and that the schools are reacting instead of being proactive.

The challenges we found in our study are similar to those reported in cyberbullying research over the past twenty years. Research on school cyberbullying policies has found several key elements that make programs effective: formal procedures for investigating incidents, clear reporting protocols, procedures for notifying parents of victims and perpetrators, established consequences, and procedures for referring students for counseling \cite{wiseman2011cyberbullying}. These findings highlight that prevention-focused approaches work better than only using punishment. They also show the importance of including the topic in the curriculum and applying rules consistently across the whole school \cite{couvillon2011recommended}. Instead of creating completely new systems for AIG-NCII, schools could adapt their current cyberbullying prevention programs to address this new problem. This approach would use twenty years of research and practical experience. It would also help schools shift from only reacting to problems after they happen to preventing them before they occur.

\subsection{Design Suggestions for Curriculum}

Education is a long-term solution because it equips students with the knowledge and values they need to act responsibly \cite{aspin2007values}. In Sections 4.2 and 4.3, educators shared their institutions' practices for addressing AIG-NCII, as well as what they think would be useful for future curriculum. However, they normally do not have adequate teaching resources specific to AIG-NCII and also lack methodologies to communicate well with students. Thus, we propose the involvement of experts in law, mental health, and technology to develop specialized materials that schools alone cannot produce.

\subsubsection{Integrating pedagogy for digital ethics and scenario-based learning.}
Digital ethics education literature highlights that adolescents benefit most from pedagogy that integrates moral reasoning, values clarification, empathy-building, and case-based reflection \cite{polizzi2022wisdom}. Online-safety research similarly emphasizes digital resilience, scenario-based learning, and opportunities for students to practice navigating dilemmas \cite{sun2022digital, lee2023developing}. Scenario-based learning avoids explicit or sensitive content but provides room to explore consent, harm, and responsibility. In Section 4.4.2, educators expressed that one of the challenges is communicating with students because adolescent students tend not to listen to teachers and try to oppose what teachers say. Media based approach such as simulations-based games or story-based games may support early ethical reflection. These interactive games could be an option but need to be done in a sensitive way because there are critiques of serious games for privacy or ethics education \cite{zhonggen2019meta}.

\subsubsection{Balancing "curiosity" with positive use of AI}

Effective AI education requires careful design that balances teaching about risks while encouraging positive engagement. Our findings in Section 4.3.2 reinforce the importance of ensuring students understand the multifaceted consequences of AIG-NCII, including social, mental health, and legal repercussions. However, educators in Section 4.4.2 cautioned that introducing AIG-NCII content within sexual education curricula might inadvertently spark curiosity and lead to misuse. Thus, we need to carefully balance pedagogical approaches that acknowledge risks while fostering critical thinking. To address this tension, we recommend that educational interventions not only emphasize the harms associated with AIG-NCII but also guide students toward constructive and ethical applications of AI technologies \cite{ma2025fostering}. By demonstrating positive use cases and creative possibilities, educators can channel students' natural curiosity toward responsible innovation rather than harmful experimentation. To support effective implementation of these educational programs, training materials should be designed to be accessible and practical for classroom use. Training modules may break down the explanation of generative AI into simple concepts with visual demonstrations rather than high-level abstract jargon. Teachers and students co-design to ensure that materials respond to classroom realities, hence easy adoption and sustainability.

\subsubsection{Establishing knowledge-sharing platforms for evolving threats.}
Educators indicated that as AI capabilities advance rapidly, school policies or responses may not be able to update in timely and effective ways to address new situations. To keep up with fast AI development, we propose building a knowledge sharing platform wherein schools and educators can access and share guidelines, experiences, and best practices. The threats from generative AI evolve rapidly, leading schools to face information gaps. Teachers and administrators act without knowing how other institutions respond to emerging risks. Schools operate in isolation when they have the same problems. A central platform would make learning faster across institutions by making visible strategies that are effective. For instance, if one school developed an effective awareness campaign or discovered a workshop template for parents, others could adopt it immediately rather than starting from zero. This helps ensure that progress in one place benefits many others.

\subsection{Parents as Digital Safety Partners}

Parents, as the closest individuals to students, play a crucial role in guiding children's online behavior and safety awareness \cite{ma2025weighing}. In the context of online safety, parental mediation includes enabling and restrictive mediation, and emphasizes that effective online-safety education requires ongoing, supportive conversations rather than restrictive warnings \cite{livingstone2017maximizing, livingstone2008parental}. Parents’ digital literacy also shapes adolescents’ vulnerability to online harms \cite{chen2025empowering}.
Thus, we can invite them to participate in co-designed workshops with experts to develop supportive lightweight training resources. The workshop can also equip them with practical skills in talking about digital risks with children. Short materials like two-minute videos that are easy to read can break down complex risks because parents have a workload. They should also maintain contact with schools to stay updated and engaged. 

\subsection{Establishing Accountability Frameworks in AIG-NCII Policies}

Based on the interviews, illustrating a role-based accountability framework can reduce confusion over who should take action in actual cases of AIG-NCII. Educators emphasized that designating clear responsibilities during policy development is critical to ensuring effective institutional responses. When institutions require these policies, it is essential to consider findings from previous research on harassment in other contexts \cite{o2004answering, choo2019metoo, schoenebeck2021youth}. They demonstrated that accountability is a critical component of effective policy implementation. Without accountability mechanisms, policies remain ineffective, regardless of whether they exist on paper. State-level laws provide essential baseline protections against AIG-NCII by creating enforceable accountability structures within educational institutions. When a student becomes a victim of AIG-NCII and the school fails to respond appropriately despite existing state protective laws, the student can pursue legal action against the institution for negligence and also they can suit the perpetrators. 

\subsection{Limitations and Future Work}

Based on our recruitment materials, participants who chose to participate may represent educators who are already aware of or concerned about AIG-NCII in schools. Therefore, our findings may not reflect the broader population of educators who have not yet encountered AIG-NCII or do not view it as an urgent issue. Additionally, this study focuses on the U.S. context and may not generalize to educational systems in other countries with different cultural norms, legal frameworks, or technological infrastructures. 
Our data rely on self-reported perceptions and recollections of past incidents. We did not verify specific cases or examine actual school policy documents. Furthermore, we captured educators' perceptions at a single point in time during a period of rapidly evolving generative AI capabilities. As AI technology continues to advance, educators' concerns and experiences may shift, emphasizing the importance of continued research in this fast-changing field.

Despite these limitations, our findings point to several important directions for future work on AIG-NCII. For a more representative understanding of the phenomenon, a larger-scale survey could capture broader aspects of educators' experiences and identify patterns across diverse educational contexts. HCI researchers can help prevent AIG-NCII by working with professionals in different fields. Workshops involving education, legal, psychological, and technological experts can be organized to subsequently train educators and parents while creating communication plans targeting students. These researchers may also spearhead the design of educational materials to be pilot tested across different regions and deposited in a central repository for schools to easily access at minimal costs. Building interdisciplinary networks and online communities allows knowledge to move instead of staying isolated. Future studies should also look at AIG-NCII in different cultures and school settings. Cultural norms change how teachers, parents, and students discuss sensitive issues. Age groups also matter: younger students may lack awareness of digital safety, while college students may face peer pressure or relationship issues. Understanding these differences can help create prevention strategies that fit both cultural and age contexts.

\section{Conclusion}

This study shows that AI-generated non-consensual intimate imagery (AIG-NCII) is an urgent issue in schools, with serious consequences for students and challenges for educators. Our findings show that many educators concerned about the negative impacts of AIG-NCII on students. They also highlight the importance of prevention, not only punishment. Schools should be involved in AI literacy, digital ethics, and consent education toward their students while also training educators on how to recognize cases and respond to them appropriately. Parents, policymakers, and technology companies must work together with schools to build stronger safeguards for adolescents. Practical steps, such as knowledge-sharing platforms and clear reporting guidelines, can help schools learn from each other and respond more quickly. By understanding educators' perspectives, this research expands HCI discussions on generative AI and offers directions for designing policies, interactive tools, and curricula that protect young people in a fast-changing technological environment.

\bibliographystyle{ACM-Reference-Format}
\bibliography{References}

\appendix
\section{Appendix: Semi-structured Interview Protocol}

This is a semi-structured interview protocol, so interviewers asked follow-up questions to gain deeper insights into participants' thoughts when needed.

\subsection*{Introduction}

Good morning/evening! How are you today? 
Thank you for your interest in contributing to our research. Today, we’ll be discussing AI, particularly focusing on AI-generated intimate images in school settings. 
There are no right or wrong answers—we want to hear your thoughts. This interview is confidential. Please feel free to express any of your opinions. 
Before we begin, can you please tell me: 

\begin{itemize}
    \item What’s your occupation?
    \begin{itemize}
        \item Can you please describe what you do?
        \item How many years have you been in this field?
    \end{itemize}
\end{itemize}

\subsection*{Background}

In the next section, I'll ask some questions about generative AI in general. 

\begin{itemize}
    \item Could you tell me what "generative AI" is?
    \begin{itemize}
        \item[--] If yes: Can you give an example?
        \item[--] If no: Ok, no problem! In our study we define generative AI as a tool that allows people to easily type in a prompt and generate things like text, images, music, and other forms of media. So for example, a person can type in a prompt such as “Create an image of a cat tap dancing” and the Generative AI tool will create that image in seconds.
    \end{itemize}

    \item Have you seen any AI-generated content before?
    \begin{itemize}
        \item Can you tell me what forms of AI-generated content have you seen?
    \end{itemize}

    \item Overall, how do you feel about AI-generated content?

    \item Have you seen any AI-generated images featuring a person?
    \begin{itemize}
        \item If yes: What was it?
        \item How did you feel when you saw it?
        \item If no: Skip to next question
    \end{itemize}

    \item Are you able to distinguish AI-generated images from real images?
      \begin{itemize}
        \item What aspects make it difficult or easy?
    \end{itemize}
\end{itemize}

\subsection*{Main Questions}

\subsubsection*{A Scenario of the AI-generated Non-consensual Intimate Image.}
\begin{itemize}
\item Ok, so we’ve talked about images that were created with generative AI, but I also wanna talk about people using AI to create intimate images. And what I mean by “Intimate” images are images that might be considered “sexy” or can be seen as more “explicit”. So, for example, in New Jersey, there was a highschool student who was caught creating sexual images of a classmate using generative AI and spreading it around school. And something similar also happened at a middle school in California, where there were five students who created and shared fake nudes of 16 eighth grade students using Generative AI to make them. With this in mind, we want to get your thoughts about the misuse of generative AI to create intimate imagery. For this situation, we are referring to “AI intimate imagery” specifically in the following questions, so please keep that in mind when answering. Not generative AI in general, but AI intimate imagery.
\end{itemize}
\subsubsection*{Existing Cases and Educators’ Attitudes.}

\begin{itemize}
    \item Have you heard about cases of AI intimate imagery in schools before?  
    \begin{itemize}
        \item If yes: If so, can you describe what happened? How did you feel? 
        \item If no: Have you heard about cases of AI intimate imagery outside of schools before?
    \end{itemize}
\end{itemize}

\subsubsection*{Impact on Students.}

\begin{itemize}
    \item Have any of your students expressed concerns related to AI intimate images? 
    \item Have any of your students experienced cases related to AI intimate images?
    \begin{itemize}
        \item How did it make you feel?
    \end{itemize}
    \item In your opinion, what are some impacts of AI intimate images on students?
    \item How do you think AI intimate images could impact students’ understanding of digital ethics? 
    \begin{itemize}
        \item (If participants don’t know what’s digital ethics, provide them DEFINITION)
        So digital ethics basically means any ethical issues with creating or using technology. For example, some digital ethical issues can be things such as privacy, information overload, internet addiction, digital divide, surveillance and robotics. 
    \end{itemize}
    \item Do you think there are certain groups of people who are more vulnerable to having AI intimate imagery created of them? 
    \begin{itemize}
        \item Why?
        \item Do you think certain genders are more vulnerable?
        \item Do you think certain age groups are more vulnerable?
    \end{itemize}
    \item What concerns do you have about students using AI to generate images in general? 
\end{itemize}

\subsubsection*{Educational Frameworks.}

\begin{itemize}
    \item Do you teach students about the ethical versus unethical use of generative AI in your curriculum?
    \begin{itemize}
        \item If yes: Would you be willing to share it with us? 
        \item If no: Is there a reason why? Is it something you want to implement in the future? Do you have any ideas on teaching AI ethical use? 
    \end{itemize}
    \item Are you aware of any guidelines your institution has when it comes to AI-generated content in general?
    \item Are you aware of any guidelines your institution has when it comes to AI intimate imagery? 
    \begin{itemize}
        \item If yes: What are they? How are these guidelines developed? 
        \item If no: In your opinion, should there be guidelines targeting AI intimate imagery in your institution?  
        \begin{itemize}
        \item If yes: 
            \begin{itemize}
        \item What are the guidelines you think there should be?
        \item What are you basing those guidelines on? [e.g Is it a personal experience, something you read online, etc.] 
        \item What challenges do you think your institution would face if they tried to implement those guidelines? 
            \end{itemize}
        \item If no: 
            \begin{itemize}
        \item Could you share your thoughts on why not?
        \item Was your opinion formed based on anything [e.g like a personal experience, something you read online, etc.]
        \item What challenges do you think your institution would face if they tried to implement new guidelines? 
            \end{itemize}
    \end{itemize}
    \end{itemize}
    \item Do you know any training educators receive regarding the ethical use of AI? 
        \begin{itemize}
        \item If yes: 
                \begin{itemize}
        \item What types of training are provided?
        \item Do you know how these trainings are developed?
        \item In your opinion, are they effective?
        
    \begin{itemize}
        \item If not, what are your suggestions to improve them?
    \end{itemize}
        \end{itemize}
        \item If no: 
        \begin{itemize}
        \item In your opinion, what training should educators receive?
        \item What challenges do you think there might be in implementing this training for educators?   
            \end{itemize}
            \end{itemize}
    \item What ways can students report AI intimate images in your institution? 
    \begin{itemize}
        \item If not, do you have any suggestions? 
    \end{itemize}
    \item What gaps do you see in current education when it comes to ethical use of AI? 
    \item How do you think education regarding AI intimate imagery should be incorporated into schools? 
    \begin{itemize}
        \item When do you think this education should begin? 
        \item Who do you think should be in charge of overseeing education regarding media literacy? [e.g. Educators, BOE, policymakers]
    \begin{itemize}
        \item What benefits do you see?
        \item What downsides do you see?
    \end{itemize}   
    \item How should educators guide students regarding the ethical use of AI-generated images? 
    \begin{itemize}
        \item  Are there any challenges?
    \end{itemize}  
    \end{itemize}   
        \item How should parents guide students regarding the ethical use of AI-generated images?
\end{itemize}

\subsubsection*{Institutional Response.}

\begin{itemize}
    \item How do you think schools should respond when AI-generated images are used inappropriately (e.g., cyberbullying, misinformation, harassment)? 
    \item Do you think that response should be different when it comes to AI intimate imagery specifically?
    \item What resources do you think schools need to better handle AI ethical issues? 
\end{itemize}

\subsubsection*{Strategies for Addressing AIG-NCII.}

\begin{itemize}
    \item What do you think is the responsibility of individuals who consume AI intimate imagery? 
    \item What do you think is the responsibility of technology companies when it comes to AI intimate imagery?
    \item Do you think tech companies should label AI-generated content? Why or why not?
    \item What role do you think the government should play in this context?
    \item Do you feel like you have a responsibility when you see AI-generated intimate content? Why or why not?
\end{itemize}

\subsection*{Closing}

This is the end of today’s interview! Today, we’ve discussed the misuse of generative AI imagery in educational settings, and your perception on the topic. 

\begin{itemize}
    \item Is there anything else you'd like to add?
    \item Do you have any questions for us?
    \item Do you know anyone we can interview?
\end{itemize}

Thank you so much for participating in our study! Have a nice day!

\end{document}